\documentclass[conference]{IEEEtran}
\IEEEoverridecommandlockouts
\usepackage{amsmath,amsfonts}
\usepackage{array}
\usepackage{multirow}
\usepackage{textcomp}
\usepackage{stfloats}
\usepackage{url}
\usepackage{verbatim}
\usepackage{graphicx}
\usepackage{cite}
\usepackage{caption} 
\usepackage{amsmath}
\usepackage{xspace}
\usepackage{fixltx2e}
\usepackage{subcaption}
\usepackage{algorithm}
\usepackage{algpseudocode}
\usepackage{enumitem}
\usepackage{xcolor}
\graphicspath{{Figure/}}
\algnewcommand{\algorithmicor}{\textbf{ or }}

\def\BibTeX{{\rm B\kern-.05em{\sc i\kern-.025em b}\kern-.08em
    T\kern-.1667em\lower.7ex\hbox{E}\kern-.125emX}}
\begin{document}

\title{Efficient Numerical Modeling of Near-Field Diffraction in ORIS-Assisted Free-Space Optical Links\\
\thanks{
    This work is supported in part by the JSPS KAKENHI (Grant 25K17607  and Grant 24K17272) and in part by the JST ASPIRE (Grant JPMJAP2345).
    } 
}

\author{\IEEEauthorblockN{
Tuan A. Hoang\IEEEauthorrefmark{1}, Phuc V. Trinh\IEEEauthorrefmark{2}, Shinya Sugiura\IEEEauthorrefmark{2}, Chuyen T. Nguyen\IEEEauthorrefmark{3}, Susumu Ishihara\IEEEauthorrefmark{4}, Thanh V. Pham\IEEEauthorrefmark{4}}
\IEEEauthorblockA{\IEEEauthorrefmark{1}Department of Engineering, Graduate School of Integrated Science and Technology, Shizuoka University, Japan.}
\IEEEauthorblockA{\IEEEauthorrefmark{2}Institute of Industrial Science, The University of Tokyo, Japan.}
\IEEEauthorblockA{\IEEEauthorrefmark{3}School of Electrical and Electronic Engineering, Hanoi University of Science and Technology, Vietnam.}
\IEEEauthorblockA{\IEEEauthorrefmark{4}College of Engineering, Academic Institute, Shizuoka University, Japan.}}









\maketitle

\begin{abstract}
This paper investigates near-field propagation in optical reconfigurable intelligent surface (ORIS)-assisted free-space optical (FSO) communication systems. Unlike conventional far-field scenarios, near-field propagation involves complex diffraction effects that hinder tractable closed-form analysis. To address this issue, a numerical framework for evaluating the optical field distribution of ORIS-assisted FSO links is proposed. Specifically, two numerical approaches are considered: direct Riemann-sum evaluation and a fast Fourier transform (FFT)-based method. Although the Riemann sum approach provides accurate field estimation, it incurs extremely high computational complexity due to the fine spatial discretization of the ORIS surface required at optical wavelengths. To improve computational efficiency, the optical-field calculation is reformulated as a convolution in the spatial-frequency domain, enabling efficient FFT-based propagation analysis. Simulation results demonstrate that the proposed FFT-based method achieves accuracy comparable to that of the Riemann-sum approach while significantly reducing computational complexity.
\end{abstract}


\section{Introduction}
Free-space optical (FSO) communication has emerged as a solution to meet the ever-increasing demand for high data rates and massive network capacity driven by modern digital applications \cite{saad2019vision}. By employing optical carriers in the infrared or visible spectrum, FSO systems can achieve fiber-like data rates without the need for physical cables \cite{khalighi2014survey}. This makes them highly attractive for a wide range of applications, including terrestrial backhaul links, inter-building communication, airborne and satellite networks \cite{kaymak2018survey, trinh2025optical}. 

Despite these advantages, FSO systems face several practical challenges that limit their widespread deployment. One of the most critical constraints is the strict requirement for a clear line-of-sight (LoS) between the transmitter and receiver. This requirement makes FSO links highly vulnerable to blockages caused by physical obstructions such as buildings, terrain, or temporary objects. 
To overcome this limitation, optical reconfigurable intelligent surfaces (ORIS) have recently been proposed as a promising solution \cite{najafi2019intelligent,jamali2021intelligent,wang2023optical}. By intelligently manipulating the phase and direction of incident optical waves, ORIS can create alternative transmission paths, thereby enabling non-line-of-sight (NLoS) communication. 

In the existing literature, most studies on ORIS-assisted FSO systems have focused on wave propagation in the intermediate-field and far-field regimes. In these regimes, analytical models remain relatively tractable and permit closed-form expressions \cite{ajam2023optical, ajam2021channel}. These models provide valuable insights into system design and optimization. However, the near-field regime has received comparatively limited attention. In this regime, the ORIS-reflected optical field cannot be accurately described by a simple linear phase approximation, since higher-order phase terms across the effective illuminated region of the ORIS must be retained. Consequently, the field depends sensitively on the ORIS size, beam footprint, propagation distance, phase profile, receiver aperture, and misalignment, which are coupled through nonlinear phase variations. This makes closed-form analytical modeling difficult, and conventional far-field approaches may no longer be directly applicable.

To address this gap, this work investigates near-field ORIS-assisted FSO propagation using numerical simulation techniques. The optical field distribution is first evaluated by directly discretizing the diffraction integral through a Riemann-sum approximation. While straightforward, this approach is computationally expensive because accurate near-field propagation modeling requires fine spatial sampling of the ORIS surface to capture rapid phase variations across the aperture. To overcome this limitation, this work further adopts a fast Fourier transform (FFT)-based propagation method, which transforms the convolution operation in the diffraction integral into multiplication in the spatial-frequency domain. This substantially reduces the computational burden while preserving numerical accuracy. Simulation results show that the direct Riemann-sum approach requires approximately one hour of computation, whereas the FFT-based method achieves comparable accuracy within only a few tens of seconds.

The remainder of this paper is organized as follows. Section II introduces the system and propagation models for the near-field ORIS-assisted FSO link. Section III describes the numerical approaches, including the Riemann-sum method and the FFT-based method, for evaluating the optical field distribution. Section IV presents simulation results and compares the performance of the two methods. Finally, Section V concludes the paper.

\section{Channel Model}
\subsection{Propagation Model}
\begin{figure}[t]
    \centering
    \includegraphics[width=0.45\textwidth]{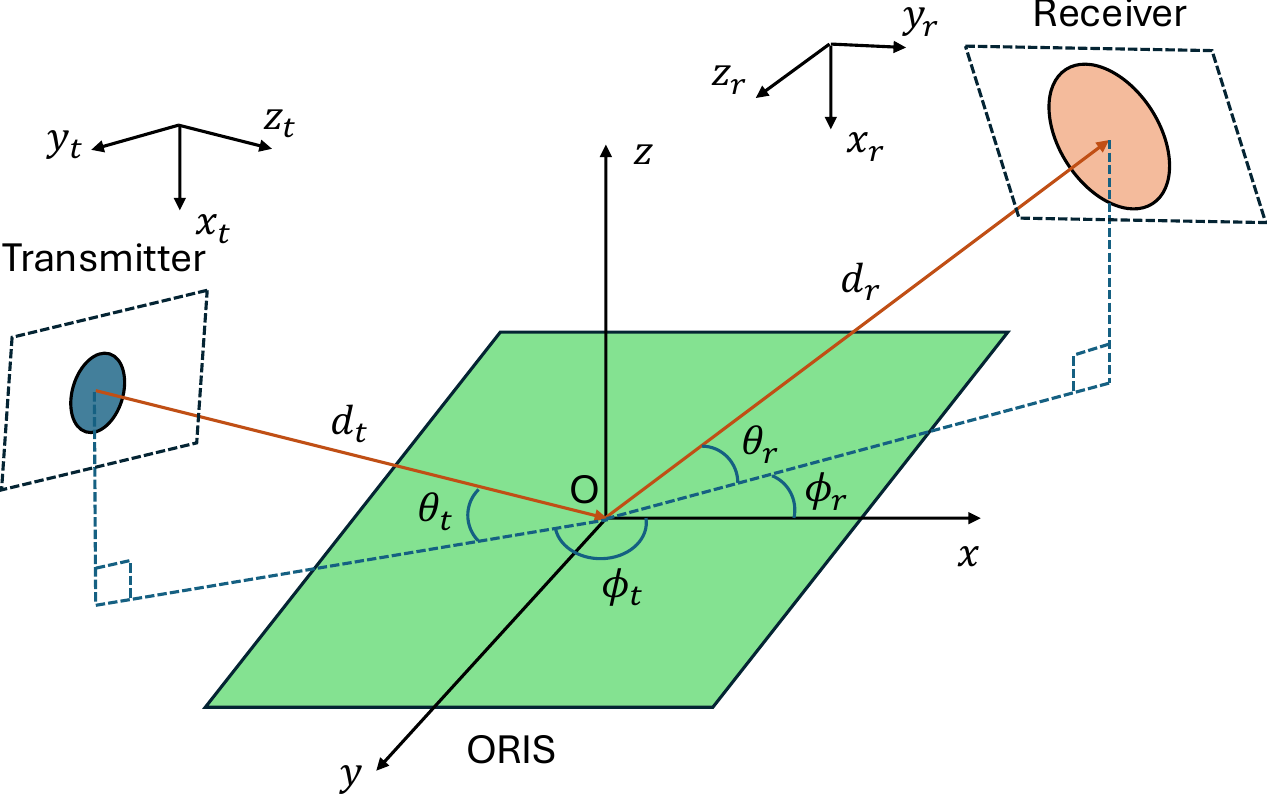}
    \caption{System configuration.}
    \label{fig:system}
\end{figure}

We consider an ORIS-assisted FSO communication system as illustrated in Fig.~\ref{fig:system}. The system consists of a transmitter, an ORIS, and a receiver. The optical beam emitted by the transmitter is reflected by the ORIS and redirected toward the receiver.
A global Cartesian coordinate system $Oxyz$ is defined with its origin being the center of the ORIS. The $Oxy$-plane coincides with the ORIS surface, and the $Oz$-axis is normal to the ORIS. In addition, a local coordinate system $O x_t y_t z_t$ is defined at the transmitter, with its origin being the center of the transmitting aperture. The $z_t$-axis is aligned with the propagation direction of the transmitted beam, while the $x_t$- and $y_t$-axes lie on the transmitter plane. Similarly, a local coordinate system $O x_r y_r z_r$ is defined at the receiver, with its origin located at the center of the receiving aperture. The $z_r$-axis is aligned with the propagation direction of the received beam, while the $x_r$- and $y_r$-axes lie on the receiver plane.
The incident and reflected beams are characterized by the elevation angles $\theta_t$ and $\theta_r$, which are defined as the angles between the beam directions and the $Oxy$ plane. The azimuth angles $\phi_t$ and $\phi_r$ are defined as the angles between the projections of the beam axes onto the $Oxy$ plane and the $Ox$ axis.

The optical field emitted by the transmitter is modeled as a Gaussian beam, which is a common laser source. The corresponding electric field in the local coordinate system is given by \cite{goodman2005introduction}
\begin{align}
&\mathbf{E}_{\mathrm{t}}(x_t,y_t,z_t)
= E_0\frac{w_0}{w(z_t)}
\exp\left(-\frac{x_t^2+y_t^2}{w^2(z_t)}\right) \nonumber \\
& \times \exp\left[-jkz_t - j\frac{k(x_t^2+y_t^2)}{2R(z_t)}
+ j\tan^{-1}\left(\frac{z_t}{z_R}\right)\right],
\end{align}
where $E_0$ denotes the field amplitude, $w_0$ is the beam waist radius, $w(z_t)$ is the beam radius at propagation distance $z_t$, $R(z_t)$ is the wavefront radius of curvature, and $j$ is the imaginary unit. These quantities are given by
$w(z_t)=w_0\sqrt{1+\left(\frac{z_t}{z_R}\right)^2}$ and
$R(z_t)=z_t\left[1+\left(\frac{z_R}{z_t}\right)^2\right]$,
where $z_R=\pi w_0^2/\lambda$ is the Rayleigh range,  $k=2\pi/\lambda$ is the wavenumber, $\lambda$ is the optical wavelength.
By transforming the Gaussian beam into the global coordinate system, the incident field on the ORIS plane can be expressed as \cite{ajam2021channel}
\begin{equation}
\mathbf{E}_{\text{ORIS}}(\mathbf{r}) = \frac{E_0 w_0}{w(\hat{d}_t)}
\exp\left(-\frac{\hat{x}^2+\hat{y}^2}{w^2(\hat{d}_t)} - j\gamma_{\mathrm{ORIS}}\right),
\end{equation}
where $\mathbf{r} = (x, y, 0)$ represents a point on the ORIS surface in the $Oxyz$ coordinate system. Here, $\hat{d}_t = d_t + x \cos(\theta_t)$ denotes the effective propagation distance accounting for oblique incidence, while $\hat{x} = x\sin(\theta_t)$ and $\hat{y} = y$ represent the projected transverse coordinates on the ORIS plane. The phase term is given by
\begin{equation}
\gamma_{\mathrm{ORIS}} = k \left( \hat{d}_t + \frac{\hat{x}^2 + \hat{y}^2}{2R(\hat{d}_t)} \right)
- \tan^{-1}\left(\frac{\hat{d}_t}{z_R}\right).
\end{equation}
To steer the reflected beam toward the receiver, the ORIS introduces a spatially varying phase shift across its surface. Specifically, a linear phase profile is applied as \cite{najafi2020physics}
\begin{equation}
\Phi_{\mathrm{ORIS}}(x,y) = k(\alpha_x x + \alpha_y y),
\end{equation}
where
\begin{align}
\alpha_x &= \cos\theta_t\cos\phi_t + \cos\theta_r\cos\phi_r, \nonumber \\
\alpha_y &= \cos\theta_t\sin\phi_t + \cos\theta_r\sin\phi_r.
\end{align}
The optical field at the receiver plane is obtained using the Huygens-Fresnel principle, which models the propagation of the reflected wave from the ORIS surface to the receiver. The received electric field is given by \cite{goodman2005introduction}
\begin{align}
\mathbf{E}_{\mathrm{r}}(\mathbf{r}_{rx})
=& \frac{1}{j\lambda}
\iint\limits_{\Sigma_{\mathrm{ORIS}}}
\mathbf{E}_{\text{ORIS}}\left(\mathbf{r})\, \exp(j\Phi_{\mathrm{ORIS}}(x, y)\right) \nonumber \\ & \times \frac{\exp\left(-jk|\mathbf{r}_{rx}-\mathbf{r}|\right)}
{|\mathbf{r}_{rx}-\mathbf{r}|}dx dy,
\label{eq:6}
\end{align}
where $\mathbf{r}_{rx} = (x_{rx}, y_{rx}, z_{rx})$ denotes a point on the receiver plane.

\subsection{Evaluation in the Far-field and Intermediate-field Regimes}
To derive a closed-form expression for the diffraction integral in \eqref{eq:6}, the previous study in \cite{ajam2021channel} proposed to use the Taylor series expansion of the propagation distance $|\mathbf{r}_{rx} - \mathbf{r}|$, expressed as follows
\begin{align}
| \mathbf{r}_{rx} - \mathbf{r} |
=& | \mathbf{r}_{rx} | 
- \frac{x x_{rx} + y y_{rx}}{| \mathbf{r}_{rx} |}
+ \frac{x^2 + y^2}{2 | \mathbf{r}_{rx} |} \nonumber \\
& - \frac{x^2 x_{rx}^2 + y^2 y_{rx}^2}{2 | \mathbf{r}_{rx} |^3}
+ \mathcal{O}\left(\frac{1}{|\mathbf{r}_{rx}|^3}\right).
\label{Taylor-expansion}
\end{align}
It has been shown in \cite{ajam2021channel} that, in the far-field regime, which satisfies the condition
\begin{equation}
\frac{x^2 + y^2}{2 | \mathbf{r}_{rx} |} \ll \lambda,
\label{eq:12}
\end{equation}
using only the first-order term, i.e., $| \mathbf{r}_{rx} | 
- ({x x_{rx} + y y_{rx}})/{| \mathbf{r}_{rx} |}$, results in a satisfactory approximation. 
To find the ORIS-receiver distance $d_r$ that satisfies \eqref{eq:12}, we first substitute the effective bounds $x_e$ and $y_e$ for $x$ and $y$. These bounds are constrained by both the ORIS aperture size $(L_x, L_y)$ and radius of the Gaussian beam footprint along the $x$- and $y$-axis, denoted as $w_x$ and $w_y$. Specifically, the effective bounds are
\[
x_e = \min\left(\frac{L_x}{2}, w_x \right), \quad
y_e = \min\left(\frac{L_y}{2}, w_y \right),
\]
Assuming that the radius of the receiver aperture is much smaller than the ORIS-receiver distance, the distance term can be approximated as $|\mathbf{r}_{rx}| \approx d_r$ for any observation point on the receiver. By substituting $x_e$ and $y_e$ into \eqref{eq:12}, we obtain the corresponding minimum far-field distance $d_f$ as
\begin{equation}
d_r \gg d_f, \quad
d_f = \frac{x_e^2 + y_e^2}{2\lambda}.
\end{equation}

As the propagation distance decreases, the quadratic term in the Taylor expansion \eqref{Taylor-expansion} must also be retained to achieve sufficient approximation accuracy. This corresponds to the intermediate-field regime, where the third-order terms remain negligible provided that
\begin{equation}
\frac{(x^2 + y^2)(x x_{rx} + y y_{rx})}{2 | \mathbf{r}_{rx} |^3} \ll \lambda,
\end{equation}
This results in the distance constraint
\begin{equation}
d_r \gg d_n, \quad
d_n = \left[ \frac{(x_e^2 + y_e^2)(x_e + y_e)}{4\lambda} \right]^{1/2}.
\end{equation}

When the propagation distance becomes comparable to or smaller than $d_n$, our preliminary study in \cite{Tuan2026} revealed that the higher-order terms in the Taylor expansion can no longer be neglected. Consequently, the diffraction integral in \eqref{eq:6} no longer admits a tractable closed-form solution and must instead be evaluated numerically.

\section{Evaluation in the Near-Field Regime}
\subsection{Riemann Sum-based Method}
\begin{figure}[t]
    \centering
    \includegraphics[width=0.39\textwidth]{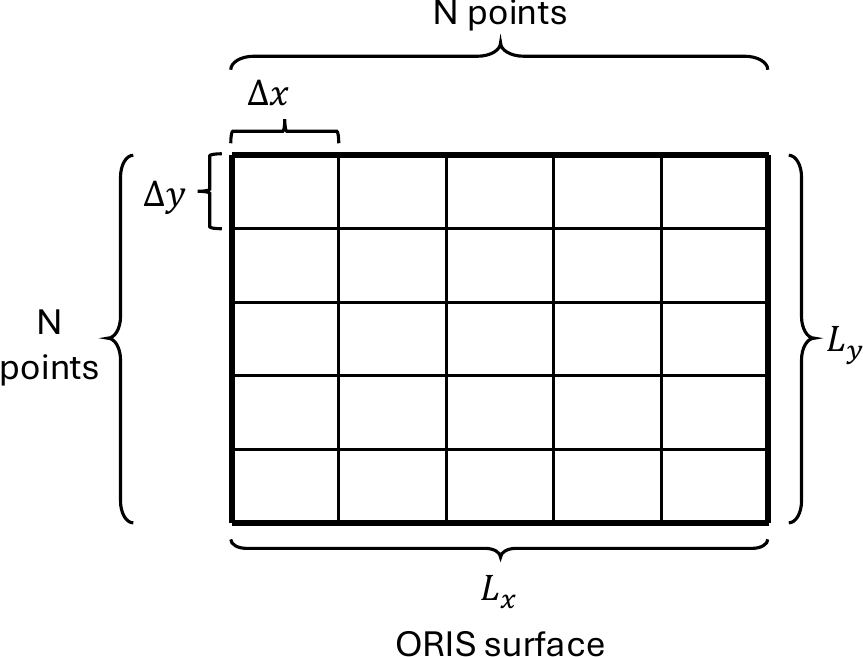}
    \caption{Discretization of the ORIS surface into 2D grids.}
    \label{fig:grid}
\end{figure}

In the near-field regime, no closed-form expression can be obtained for the electric field in \eqref{eq:6}, since higher-order terms in the phase expansion become non-negligible. A straightforward approach to evaluate the field is to approximate the integral using a Riemann summation. Specifically, the ORIS surface is discretized into a two-dimensional grid of sufficiently small elements, as illustrated in Fig.~\ref{fig:grid}. The total received electrical field is obtained by summing the contributions from all grid points
\begin{align}
&\mathbf{E}_{\mathrm{r}}(\mathbf{r}_{rx})
\approx
\frac{1}{j\lambda}
\sum_{p} \sum_{q}
\mathbf{E}_\mathrm{ORIS}(\mathbf{r}_{p,q}) \,  \nonumber \\
&\times
\text{exp}(j\Phi_{\mathrm{ORIS}}(\mathbf{r}_{p,q})) \frac{\exp\left(-jk|\mathbf{r}_{rx}-\mathbf{r}_{p,q}|\right)}
{|\mathbf{r}_{rx}-\mathbf{r}_{p,q}|}
\, \Delta x \, \Delta y,
\end{align}
where $\mathbf{r}_{p,q}$ denotes the position of the $(p,q)$-th grid element on the ORIS surface, while $\Delta x$ and $\Delta y$ represent the discretization intervals along the $x$- and $y$-axes, respectively. The number of sampling points along the $x$- and $y$-axes is denoted by $N$.

To accurately capture the rapidly varying phase of optical waves, the spatial sampling interval must be on the order of, or smaller than, the optical wavelength. As a result, for high-frequency optical signals and large ORIS apertures, an extremely fine discretization is required, resulting in a large number of sampling points and, consequently, a high computational cost.
To quantify the convergence behavior of the Riemann-sum approximation, we define the relative error between two discretization levels with different spatial resolutions. Let $\mathbf{E}_{\textrm{r}}^{(N)}$ denote the computed electric field obtained using $N$ sampling points on each spatial. The relative error between two spatial resolutions $N_i$ and $N_{i-1}$ is defined as
\begin{equation}
\epsilon_i = \frac{\left| \mathbf{E}_{\textrm{r}}^{(N_i)} - \mathbf{E}_{\textrm{r}}^{(N_{i-1})} \right|}{\left| \mathbf{E}_{\textrm{r}}^{(N_{i-1})} \right|},
\end{equation}
This metric measures the change in the numerical solution as the grid is refined and serves as an indicator of convergence.

\begin{figure}[bp]
    \centering
    \includegraphics[width=0.4\textwidth, height=0.26\textwidth]{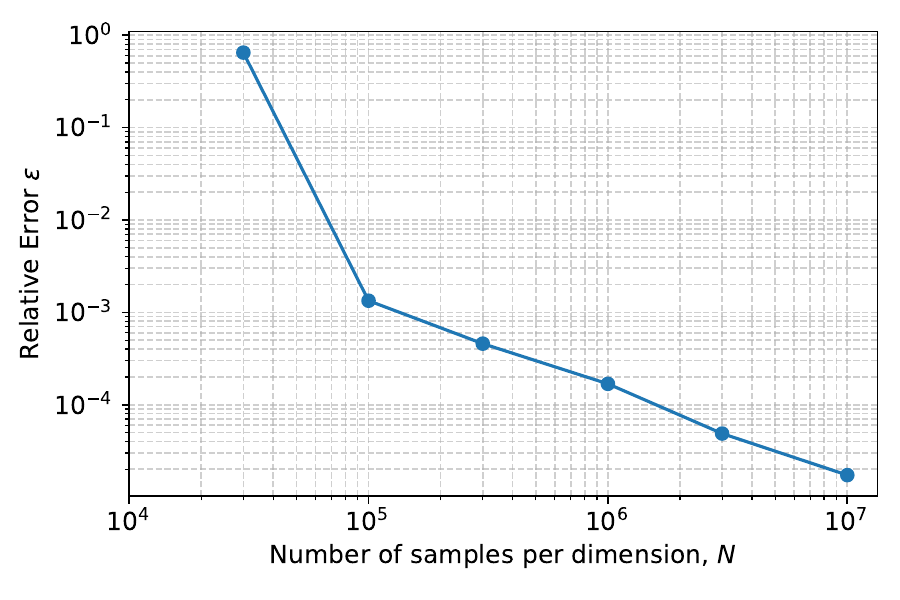}
    \caption{Convergence of the received electric field at the receiver center with respect to spatial discretization resolution.}
    \label{fig:convergence}
\end{figure}

Fig.~\ref{fig:convergence} illustrates the convergence behavior of the received electric field at the center of the receiver as the spatial resolution increases.
As observed in Fig.~\ref{fig:convergence}, the relative error decreases as the spatial resolution becomes finer, confirming the convergence of the Riemann-sum approximation. In particular, for an ORIS with size $1\text{m}\times1\text{m}$, approximately $N = 10^6$ sampling points on each dimension (corresponding to $\Delta x = \Delta y = 10^{-6}$~m) are required to achieve a relative error of less than $10^{-3}$. 
However, such a fine discretization results in prohibitively high computational complexity. Therefore, the Riemann sum-based approach is impractical for large-scale system analysis, motivating the development of more computationally efficient numerical methods.

\subsection{Fast Fourier Transform (FFT)-based Method}
To overcome the high computational cost of the Riemann-sum approach, we exploit the fact that free-space propagation can be interpreted as a convolution between the ORIS aperture field and the propagation kernel. By reformulating the propagation process in the spatial-frequency domain, the received electric field can be efficiently evaluated using FFT-based techniques \cite{filipovich2024torchoptics}, thereby significantly reducing computational complexity while maintaining high numerical accuracy.
Starting from \eqref{eq:6}, the electric field on an observation plane parallel to the ORIS and separated by a distance $d_z$ can be written in Cartesian coordinates as
\begin{align}
&\mathbf{E}_{\mathrm{r}}(x_{rx},y_{rx})
= \frac{1}{j\lambda}
\iint\limits_{\Sigma_{\mathrm{ORIS}}}
\mathbf{E}_\mathrm{ORIS}(x,y)\, \exp(j\Phi_{\mathrm{ORIS}}(x,y)) \nonumber \\
& \times
\frac{\exp\left(-jk\sqrt{(x_{rx}-x)^2 + (y_{rx}-y)^2 + d_z^2}\right)}
{\sqrt{(x_{rx}-x)^2 + (y_{rx}-y)^2 + d_z^2}} \, dx\, dy.
\label{eq:15}
\end{align}
We define the equivalent aperture field as
\begin{equation}
\psi(x,y) = \mathbf{E}_\mathrm{ORIS}(x,y)\, \exp(j\Phi_{\mathrm{ORIS}}(x,y)),
\end{equation}
and the propagation kernel as 

\begin{align}
h(x_{rx}&-x,y_{rx}-y) =
\frac{1}{j\lambda} \nonumber \\
&\times \frac{\exp\left(-jk\sqrt{(x_{rx}-x)^2 + (y_{rx}-y)^2 + d_z^2}\right)}{\sqrt{(x_{rx}-x)^2 + (y_{rx}-y)^2 + d_z^2}}.
\end{align}
With these definitions, the integral in \eqref{eq:15} can be interpreted as a two-dimensional convolution between $\psi(x,y)$ and $h(x_{rx}-x,y_{rx}-y)$, expressed by
\begin{equation}
\mathbf{E}_{\textrm{r}}(x_{rx},y_{rx}) = (\psi * h) \ (x_{rx},y_{rx}),
\end{equation}
where $*$ denotes convolution.
According to convolution theory, convolution in the spatial domain corresponds to multiplication in the spatial-frequency domain. Therefore, by applying the angular spectrum formulation, the propagated optical field at the observation plane can be expressed as
\begin{equation}
\psi_{rx}(x_{rx},y_{rx}) =
\mathcal{F}^{-1}\left\{
H_z(k_x,k_y)\,\mathcal{F}\{\psi(x,y)\}
\right\},
\end{equation}
where $\mathcal{F}$ and $\mathcal{F}^{-1}$ denote the two-dimensional Fourier transform and its inverse, respectively, and $\psi_{rx}(r_{rx})$ represents the propagated field at the receiver plane.
The transfer function $H_z(k_x,k_y)$ is given by \cite{filipovich2024torchoptics}
\begin{equation}
H_z(k_x,k_y) =
\exp\left(- j d_z \sqrt{k^2 - k_x^2 - k_y^2}\right),
\end{equation}
where $k_x$ and $k_y$ denote the spatial-frequency components along the $x$- and $y$-axes, respectively. 

When the received field is computed over an $N\times N$ observation plane from an $N\times N$ ORIS aperture grid, the direct Riemann sum-based method requires $O(N^4)$ operations. This is because each of the $N^2$ observation points must sum the contributions 
from all $N^2$ aperture samples. In contrast, the FFT-based angular-spectrum method uses the convolution property of free-space 
propagation and computes the field over the entire observation plane at once, with complexity $O(N^2\log N^2)$. Therefore, the FFT-based method is much more efficient for high-resolution simulations. However, it also requires storing the field over the whole discretized 
plane, resulting in excessive memory requirements for very large $N$.

\section{Simulation Results and Discussions}
In this section, simulation results are presented to demonstrate the computation of the electric field using the two proposed methods. Without otherwise noted, simulation parameters are provided in Table~\ref{table:parameter}.  The Riemann sum-based method uses a spatial resolution of $10^6 \times 10^6$, whereas the FFT-based method employs a $1024 \times 1024$-point FFT. 
\begin{table}[tbp]
    \centering
    \caption{Simulation parameters.}
    \label{tab:system_parameters}
    \small
    \renewcommand{\arraystretch}{1.2}
    \begin{tabular}{l c l}
\hline\hline
\textbf{Parameter} & \textbf{Symbol} & \textbf{Value} \\ \hline\hline

Optical wavelength & $\lambda$ & 1550 nm \\ \hline
Electric field at the origin & $\mathbf{E}_{0}$ & 1 V/m \\ \hline
ORIS size & $L_{x} \times L_{y}$ & $1 \,\text{m} \times 1 \,\text{m}$ \\ \hline
ORIS-transmitter distance  & $d_{t}$ & 1 km \\ \hline
ORIS-receiver distance & $d_r$ & 5 m \\ \hline
Orientation of the transmitter & $(\theta_t,\phi_t)$ & $(\frac{\pi}{3}, 0)$ \\ \hline
Orientation of the receiver & $(\theta_r,\phi_r)$ & $(\frac{\pi}{4}, \pi)$ \\ \hline
\label{table:parameter}
    \end{tabular}
\vspace{-0.3cm}
\end{table}

\begin{figure}[ht]
    \centering
    \includegraphics[width=0.4\textwidth]{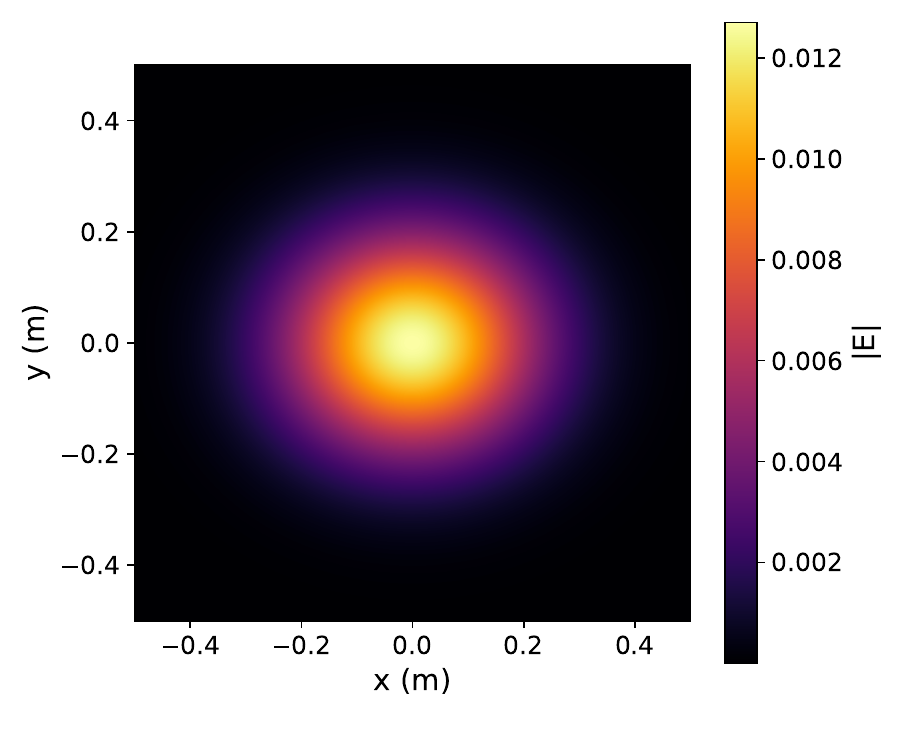}
    \caption{Amplitude distribution of the electric field on the observation plane.}
    \label{fig:observation plane}
\end{figure}

Firstly, Fig.~\ref{fig:observation plane} illustrates the amplitude distribution of the electric field at the observation plane located at $d_z = d_r \sin(\theta_r) \approx 3.53$m. Although the distribution follows a Gaussian-like shape as expected, the peak intensity is located at the center of the $Oxy$ plane instead of coinciding with the receiver center.
This discrepancy stems from the small number of FFT points (i.e., $1024 \times 1024$ points) used in simulations, which introduces inaccuracies in the phase representation. While it can be resolved by increasing the number of FFT points, it requires significantly more computational memory. This misalignment, however, is not a serious issue since the direction of the reflected beam from the ORIS is characterized by known parameters  $\theta_r$ and $\phi_r$, and $d_r$. Hence, the observation plane can be shifted to its correct spatial location along the $x$- and $y$-axes by
$
{d_r}{\cos(\theta_r)} \cos(\phi_r) \quad \text{and} \quad {d_r}{\cos(\theta_r)} \sin(\phi_r),
$
respectively.

Fig.~5 compares the electric field amplitude distributions on the receiver surface along $x_r$-axis obtained using the Riemann-sum-based and FFT-based methods for two beam waist radius, $w_0 = 2.5\,\text{mm}$ and $w_0 = 5.0\,\text{mm}$. In both cases, the FFT-based approach produces distributions that closely match those obtained with the Riemann sum-based implementation, demonstrating strong agreement between the two methods.

In Fig.~6, the relative error of the FFT-based method compared to the Riemann sum-based method is shown. The error remains small near the center of the receiver and gradually increases toward the edge regions. Nevertheless, because most of the optical energy is concentrated near the center of the beam footprint, errors in the peripheral region have a negligible effect on the overall propagation characteristics. Our simulations also verify that the FFT-based method significantly reduces computation time. Specifically, while the Riemann sum-based approach requires approximately one hour to complete the calculation, the FFT-based implementation achieves the result in an average of only 13.1 seconds using an NVIDIA RTX A4000 GPU with parallelized Python code. 
\begin{figure}[tbp]
    \centering
    \includegraphics[width=0.45\textwidth,height=0.3\textwidth]{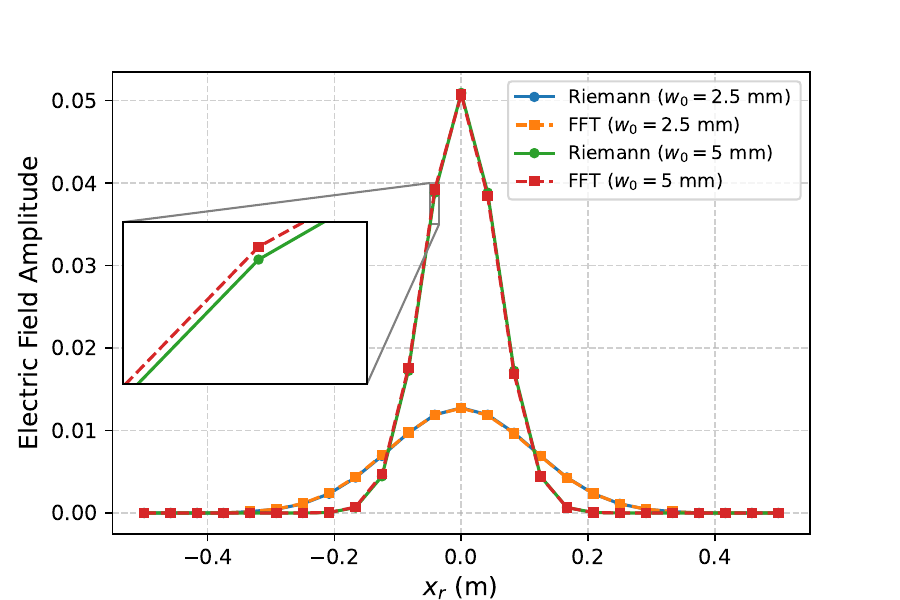}
    \caption{Comparison of electric field amplitude distributions between the Riemann sum-based and FFT-based methods.}
    \label{fig:compare}
\end{figure}
\begin{figure}[ht]
    \centering
    \includegraphics[width=0.45\textwidth,height=0.3\textwidth]{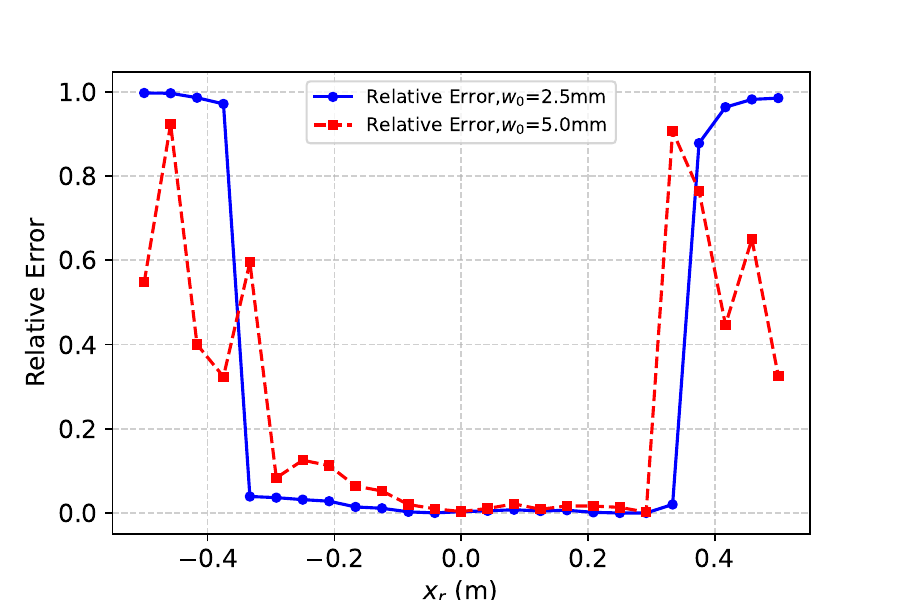}
    \caption{Relative error of the FFT-based method compared to the Riemann sum-based method.}
    \label{fig:error}
\end{figure}

\section{Conclusion}
This paper presented a numerical framework for near-field optical propagation in ORIS-assisted FSO systems. Although the Riemann sum-based approach provides high numerical accuracy, it incurs substantial computational complexity due to the extremely fine spatial discretization of the ORIS surface. In contrast, the FFT-based method significantly reduces the computational burden by performing the propagation in the spatial-frequency domain. Simulation results demonstrated that the FFT-based method accurately reproduces the electric-field distribution obtained using the Riemann-sum-based approach while achieving a significantly lower computational cost.

\bibliographystyle{ieeetr}   
\bibliography{reference} 

\end{document}